\documentclass[pre,twocolumn,a4paper,showpacs,showkeys]{revtex4-1}

\usepackage[utf8]{inputenc}
\usepackage{amsmath,amsfonts,amssymb}
\usepackage{graphicx,graphics}
\usepackage{color,textcomp,bm}
\usepackage{subfigure,array,xspace} 
\usepackage[squaren,Gray,cdot]{SIunits}

\usepackage[pdfstartview=FitV,bookmarks=true, linktocpage=true,colorlinks=true, urlcolor=blue,linkcolor=red, citecolor=blue]{hyperref}   
\newcommand{\ie}{\textit{i.e.}\@\xspace}

\newcommand{\etal}{\emph{et al.~}}

\DeclareMathOperator{\Tr}{Tr}

\usepackage{fancyhdr} 

\begin{document}

\title{Negative reflection of Lamb waves at a free edge:\\ tunable focusing and mimicking phase conjugation}

\author{Benoît Gérardin}
\author{Jérôme Laurent}
\author{Claire Prada}    \email{claire.prada@espci.fr}
\author{Alexandre Aubry} \email{alexandre.aubry@espci.fr}
\affiliation{ESPCI ParisTech, PSL Research University, CNRS, Institut Langevin, UMR 7587, 1 rue Jussieu, F-75005 Paris, France}
\date{\today}

\begin{abstract}

The paper studies the interaction of Lamb waves with the free edge of a plate. The reflection coefficients of a Lamb mode at a plate free edge are calculated using a semi-analytical method, as a function of frequency and angle of incidence. The conversion between forward and backward Lamb modes is thoroughly investigated. It is shown that, at the zero-group velocity (ZGV) frequency, the forward $S_1$ Lamb mode fully converts into the backward $S_{2b}$ Lamb mode at normal incidence. Besides, this conversion is very efficient over most of the angular spectrum and remains dominant at frequencies just above the ZGV-point. This effect is observed experimentally on a Duralumin plate. Firstly, the $S_1$ Lamb mode is selectively generated using a transducer array, secondly the $S_{2b}$ mode is excited using a single circular transducer. The normal displacement field is probed with an interferometer. The free edge is shown to retro-focus the incident wave at different depths depending on the wave number mismatch between the forward and backward propagating modes. In the vicinity of the ZGV-point, wave numbers coincide and the wave is retro-reflected on the source. In this frequency range, the free edge acts as a perfect phase conjugating mirror.
\end{abstract}



\maketitle

\section{Introduction}

Controlling the propagation of acoustic or elastic waves is of fundamental interest for many applications ranging from imaging the living and detecting hazardous components, to information processing and structural health monitoring. Owing to this context, the concept of negative refraction has received a great deal of attention for the last fifteen years. In a negative index material, the energy flow as dictated by the Poynting vector should be in the opposite direction to the wave vector~\citep{veselago1968electrodynamics,pendry2000negative,pendry2004negative}. This property implies that at an interface between a positive and a negative index material sound is bent the unusual way relative to the normal. In this paper, we consider a related phenomenon referred to as negative reflection. A negative reflecting mirror is an interface at which sound is retro-reflected. There is a strong analogy with a phase conjugating mirror~\cite{pendry2008time}, except that, in our case, the negative reflecting mirror does not involve any nonlinear process.\\

Until now, negative refraction of elastic waves has been mainly implemented either using the concept of phononic crystals~\citep{morvan2010experimental,croenne2011negative,pierre2010negative,dubois2013flat}, or opting for metamaterials~\citep{zhu2014negative}, an arrangement of tailored subwavelength building blocks from which the material gains unusual macroscopic properties. However such manmade materials usually rely on resonating structures, a feature that induces strong losses. More recently, an alternative route has been proposed for the negative refraction of guided elastic waves. An elastic plate actually support an ensemble of modes, the so-called Lamb waves, which exhibit complex dispersion properties. In particular, some Lamb modes, often referred to as backward propagating modes in the literature, naturally display a negative phase velocity~\cite{tolstoy1957wave,mindlin1960waves,meitzler1965backward,negishi1987existence}. These modes originate from the repulsion between two dispersion branches having close cut-off frequencies, corresponding to a longitudinal and a transverse thickness mode of the same symmetry. The lowest branch exhibits a minimum corresponding to a zero-group velocity (ZGV) point~\citep{prada2005laser,holland2003air}. Above this resonance, there is a coexistence of a negative phase velocity (backward) mode and a positive phase velocity (forward) mode. This peculiar property has been taken advantage of to achieve negative refraction~\citep{bramhavar2011negative,philippe2015focusing} through mode conversion between forward and backward propagating modes (or vice versa) at a step-like thickness discontinuity. In a pionneering experiment, few years ago, Germano \etal showed that this conversion from forward to backward Lamb waves also occurs at a simple free edge of an elastic plate, giving rise to negative reflection~\cite{germano2012anomalous}. More recently, in a concurrent study  by Veres \etal, the negative reflection of Lamb waves has been investigated above the ZGV resonance through numerical simulations and laser-ultrasound experiments~\cite{veres2016broad}.\\

In this paper, we investigate this phenomenon in-depth both theoretically and experimentally. From the theoretical side, we compute the reflection coefficient of a Lamb mode impinging on the free edge of the plate as a function of frequency and angle of incidence. Several studies addressed this problem in the past under normal~\citep{torvik1967reflection,auld1977variational,gregory1983reflection,morvan2003lamb,galan2002numerical,pagneux2006revisiting} and oblique~\cite{gunawan2007reflection,wilcox2010scattering,santhanam2013reflection,feng2016scattering} incidence but none of them considered the case of backward Lamb modes. Our main result is that, close to the zero-group velocity point, the forward and backward modes are strongly coupled. They convert preferentially into each other, giving rise to negative reflection over the whole angular spectrum.  These theoretical results are then illustrated through an experiment conducted on a Duralumin plate.  The incident mode is selectively generated by a transducer array in contact, and the field reflected at the plate edge is probed by laser interferometry. Whereas the pioneering study of Germano \etal~\cite{germano2012anomalous} only dealt with the negative reflection of plane waves, a cylindrical incident wave-field is here considered.  The free edge is shown to back-focus the incident wave-field at different depths according to the wave number mismatch between the forward and backward propagating modes. In the vicinity of the ZGV-point, this mismatch vanishes and the reflected wave-field back-focuses exactly on the point-source: The free edge then acts as a phase conjugating mirror.

\section{Theoretical investigation of the negative reflection of Lamb waves at a free edge}\label{sec2}

The interaction of an incident Lamb wave with the free edge of a semi-infinite plate has been widely studied under normal incidence, using modal decomposition and collocation techniques~\citep{torvik1967reflection,auld1977variational,gregory1983reflection,morvan2003lamb,galan2002numerical,pagneux2006revisiting}. The case of an oblique incidence has been considered more recently. Gunawan \etal~\cite{gunawan2007reflection} and Wilcox \etal~\cite{wilcox2010scattering} addressed this problem in the low frequency regime using a modal decomposition and a frequency-domain finite element method, respectively. More recently, Santanham \etal~\cite{santhanam2013reflection}, and Feng \etal~\cite{feng2016scattering} have investigated the reflection  of Lamb modes in a higher frequency regime, using a modal decomposition. Nevertheless, in the above mentionned references, the question of backward modes is not addressed as they consider frequency bands over which backward propagation does not occur. In this paper, we consider the reflection of a Lamb wave at the free edge of a plate in the vicinity of a ZGV resonance at which forward and backward propagating modes coexist. To that aim, we extend the modal decomposition method derived by Pagneux~\cite{pagneux2006revisiting} under normal incidence to a 3D polarization.  

\subsection{Determination of the plate modes}\label{sec2a}

The plate modes are first derived in the right-handed coordinate system $\left(x_1,x_2,x_3\right)$ whose  $x_1-$axis lies along the propagation direction of the wave and $x_2-$axis along the normal of the plate.  Considering a homogeneous, isotropic plate at the pulsation $\omega$, the elasticity equations are given by  
\begin{equation}
-\rho \omega^2 \mathbf{u} = \mathbf{\nabla} \cdot \boldsymbol{\sigma},
\label{eq:elasticity}
\end{equation}
where $\rho$ is the density of the material, $\mathbf{u} = \left(u_1,u_2,u_3\right)^T$ is the displacement field and $\boldsymbol{\sigma} = \left[\sigma_{ij}\right]$ is the stress tensor. Considering an isotropic material, the strain-stress relations can be written as $\sigma_{ij} = \lambda \Tr{\left[\boldsymbol{\epsilon}\right]} \delta_{ij} + 2 \mu \epsilon_{ij}$, with $\left(\lambda,\mu\right)$ the Lam\'e parameters, $\delta_{ij}$ the Kronecker symbol and $\boldsymbol{\epsilon} = \left[\epsilon_{ij}\right]$ the infinitesimal strain tensor. The boundary conditions correspond to the cancellation of the stress tensor on the plate surfaces, $\boldsymbol{\sigma} \cdot \mathbf{n} = \mathbf{0}$, where $\mathbf{n}$ is the normal to the surface boundary. Considering the geometry of the problem, solutions are in the form :
\begin{equation}
\left\lbrace u_i \left(x_1,x_2\right),\sigma_{ij}\left(x_1,x_2\right)\right\rbrace = \left\lbrace u_i \left(x_2\right),\sigma_{ij}\left(x_2\right)\right\rbrace \cdot \exp{\left(\imath \cdot k x_1\right)}.
\end{equation}
Dimensionless variables are obtained by normalizing the components of the displacement field $\mathbf{u}$ by $h$, the coordinates $x_i$ by $h$ and the stress tensor $\boldsymbol{\sigma}$ by $\mu$. The resulting equations are 
\begin{eqnarray}
-\Omega^2 \mathbf{u} & = & \mathbf{\nabla} \cdot \boldsymbol{\sigma},\\	\label{eq:elasticity_v1}
\sigma_{ij} & = & \left( \gamma -2 \right) \Tr{\left[\boldsymbol{\epsilon}\right]}  \delta_{ij} + 2\epsilon_{ij},\\ \label{eq:elasticity_v2}
\boldsymbol{\sigma}\left(x_2 = \pm 1\right)\cdot\mathbf{n_2} & = & 0,	\label{eq:elasticity_v4}
\end{eqnarray}
with $\Omega = \omega h/c_T$, the dimensionless frequency, $\gamma = (\lambda+2\mu)/\mu = c_L^2/c_T^2$, $c_L$ and $c_T$, the bulk longitudinal and shear velocities, and $\mathbf{n_2}$ the normal to the plane $\left(x_1,x_3\right)$.\\

In this system the equations associated with the different displacement components are uncoupled. This decoupling distinguishes two different families of modes, shear horizontal (SH) waves with polarization $u_3$ and Lamb waves, polarized in the propagation plane $\left(u_1,u_2\right)$. Both families are composed of an infinite number of modes called propagative for a real wave number, evanescent for an imaginary wave-number  and inhomogeneous for a complex wave number. At a given frequency, there is only a finite number of propagating SH and Lamb modes, whereas an infinite number of evanescent SH modes and inhomogeneous or evanescent Lamb modes exists. Considering the symmetries of the displacement field with respect to the plane $x_2 = 0$, both Lamb and SH modes can be separated in two independent subfamilies of symmetrical and antisymmetrical modes. Symmetric/antisymmetric SH modes correspond to an even/odd $u_3\left(x_2\right)$ polarization along the plate thickness. Symmetric/antisymmetric Lamb modes have an even/odd in-plane component $u_1\left(x_2\right)$ combined with an odd/even transverse component $u_2\left(x_2\right)$. As the plate edge is symmetrical with respect to the $x_2=0$ plane, the scattering at a free edge preserves the mode's symmetry.  As a consequence, a symmetrical Lamb mode is reflected into symmetrical Lamb and SH modes.\\

Right-going (respectively left-going) propagating modes correspond to a positive (respectively negative) group velocity $\partial \omega/\partial k$, whereas right-going (respectively left-going) evanescent and inhomogeneous modes correspond to wave numbers with strictly positive (respectively negative) imaginary parts. In the following, each (Lamb and SH)  right-going mode, $\left\lbrace {u}_i^{\left(n\right)},{\sigma}_{ij}^{\left(n\right)}\right\rbrace$, is associated with an index $n$. These modes are ranged in ascending order of their wave numbers imaginary part and descending order of their real part. Each right-going mode, $\left\lbrace {u}_i^{\left(n\right)},{\sigma}_{ij}^{\left(n\right)}\right\rbrace$, of wave number $k_n$ can be associated with a left going mode, $\left\lbrace \tilde{u}_i^{\left(n\right)},\tilde{\sigma}_{ij}^{\left(n\right)}\right\rbrace$, of wave number $-k_n$.

\begin{figure}[!ht]
\centering
\includegraphics[width=0.9\columnwidth]{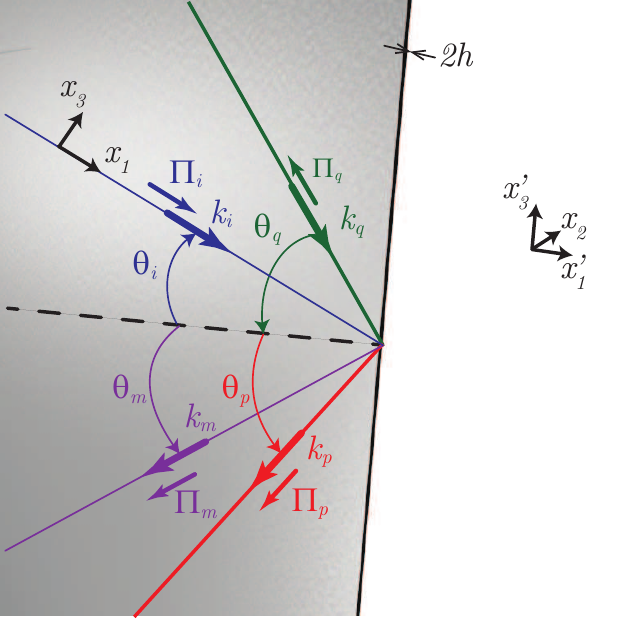}
\caption{(color online) Interaction of an obliquely incident Lamb wave with the free edge of a semi-infinite plate. To satisfy the stress-free condition at the free edge, the monochromatic incident wave is reflected into an infinite combination of the propagating, inhomogeneous, evanescent Lamb and SH modes existing at this frequency. Each mode is reflected with an angle $\theta_{n}$ dictated by the conservation of the wave-vector component along $x_3$. Here we show the case of an incident mode $i$ that is positively reflected into two forward modes, $m$ and $p$, and negatively reflected into a backward mode $q$. For the sake of clarity, evanscent and inhomogeneous modes are not represented in the figure but note that they are taken into account in our calculation. Note that the Poynting vector $\Pi_q$ of the negatively reflected mode $q$ is anti-parralel to its wave vector $k_q$.}
\label{fig1}
\end{figure}

\subsection{Interaction of a Lamb mode with the free edge of a plate}\label{sec2b}

The plate is now assumed to be semi-infinite, occupying the region $x^{\prime}_1 < 0$, bounded by two parallel planes at $x_2 = \pm h$, as depicted in Fig.~\ref{fig1}. Let us consider an incident right-going Lamb mode, of wave number $k_i$, carrying an unit energy flux towards the free edge of the plate with an angle of incidence $\theta_i$ with respect to $\mathbf{n^{\prime}_1}$, the normal to the free edge, as depicted in Fig.~\ref{fig1}. In order to satisfy the stress-free condition, this incident Lamb mode is reflected into an infinite combination of left-going Lamb and SH modes of wave numbers $-k_n$. Thus, it is necessary to consider not only the propagating modes but also the different evanescent and inhomogeneous Lamb and SH modes. For a given mode $n$, the reflection angles $\theta_{n}$ are determined by the conservation of the component $\xi$ of the wave vector along the edge direction. 
\begin{equation}
\xi = k_i \sin{\left(\theta_i\right)} = -k_n \sin{\left(\theta_{n}\right)}.
\label{eq:angle_reflection}
\end{equation}
Note that for an incident forward mode associated with a positive angle, the reflection angle is positive for a forward mode and negative for a backward mode. The strain-displacement field $\left\lbrace u_i^{\prime\left(n\right)},\sigma_{ij}^{\prime \left(n\right)}\right\rbrace$ of a reflected mode expressed in the coordinate system $\left(x^{\prime}_1,x_2,x_3^{\prime}\right)$, associated with the plate edge, can be obtained from the strain-displacement field $\left\lbrace {u}_i^{\left(n\right)},{\sigma}_{ij}^{\left(n\right)}\right\rbrace$ expressed in the coordinate system $\left(x_1,x_2,x_3\right)$ using the following equations :
\begin{eqnarray}
\mathbf{u^{\prime\left(n\right)}} &=& R\left(\theta_{n}\right) \cdot \mathbf{u^{\left(n\right)}}, \label{u_rotation}\\
\boldsymbol{\sigma^{\prime\left(n\right)}} &=& R\left(\theta_{n}\right) \cdot \boldsymbol{\sigma^{\left(n\right)}} \cdot R\left(\theta_{n}\right)^T, \label{sigma_rotation}
\end{eqnarray}
where $R\left(\theta\right)$ is the rotation matrix,
$$R\left(\theta\right) = 
\left[ \begin{array}{ccc} 
\cos{\left(\theta\right)}& 0 &-\sin{\left(\theta\right)}\\
0& 1 &0\\
\sin{\left(\theta\right)}& 0 &\cos{\left(\theta\right)}
\end{array}\right].
$$
The stress-free condition at the edge of the plate ($x^{\prime}_1=0$) can be written as the cancellation of the sum of the incident mode and the reflected modes weighted by their complex reflection coefficients $r_{i \vert n}$ :
\begin{equation}
\mathbf{0} = \boldsymbol{{\sigma}^{\prime\left(i\right)}} \cdot \mathbf{n_1} + \sum_{n=1}^{+\infty} r_{i \vert n} \boldsymbol{\tilde{\sigma}^{\prime\left(n\right)}}\cdot \mathbf{n_1}.
\label{eq:boundary_condition}
\end{equation}
The solution of this problem is calculated following the method developed by Pagneux~\cite{pagneux2006revisiting} and extending it to a 3D polarization. It combines a collocation discretization along the $x_2$-coordinate and a modal approach. This approach, fully described in the Appendix, allows to derive the amplitude reflection coefficient $r_{i\vert n}$ for any incoming mode and incident angle $\theta_i$. Note that, due to reciprocity, $r_{i\vert n}= r_{n\vert i}$. In practice, in the frequency band starting from $f_{ZGV} \times\left(2h\right) = 2.87~\textrm{MHz}\cdot\textrm{mm}$ to $f\times\left(2h\right) = 3.30~\textrm{MHz}\cdot\textrm{mm}$, we have considered 199 Lamb modes and 100 SH modes, which allows to satisfy the conservation of energy, since  $1 - \sum_{p} \left| r_{i\vert p} \right|^2 \sim 10^{-8}$, where the index p denotes a sum over the propagating modes.

\subsection{Application to a Duralumin plate}\label{sec2c}

\begin{figure*}[ht!]
\centering
\begin{minipage}[c]{0.7\textwidth}
\subfigure{\includegraphics[width=\columnwidth]{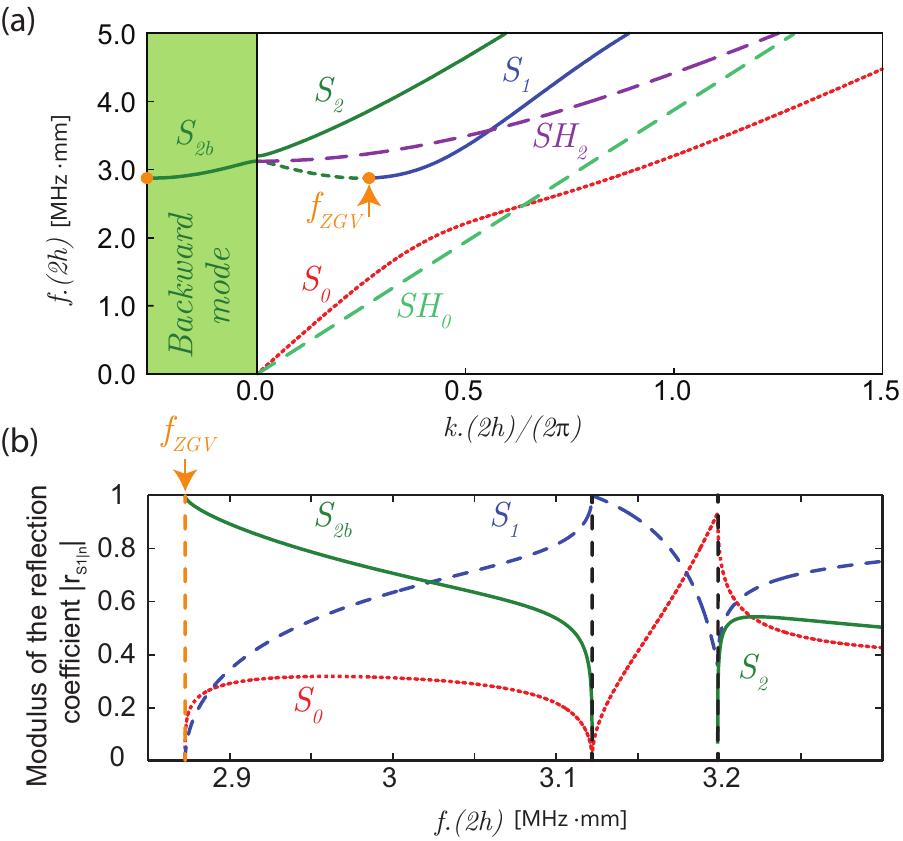}\label{fig2a}}
\subfigure{\label{fig2b}}
\end{minipage}\hfill
\begin{minipage}[c]{0.25\textwidth}
\caption{(color online)  (a) Dispersion curves of the first propagating symmetrical Lamb and SH modes in a Duralumin plate. (b) Modulus of the reflection coefficients $\left| r_{S_1,n} \right|$ of the propagating Lamb modes for an incoming $S_1$ Lamb mode at normal incidence as function of the product (frequency$\cdot$thickness) in MHz.mm.}
\label{fig2}
\end{minipage}
\end{figure*}

Our theoretical approach is now applied to the case of a Duralumin plate, with a density $\rho = 2790~\textrm{kg/m}^3$, a  longitudinal wave velocity $c_L = 6398~\textrm{mm/}\mu\textrm{s}$, and a transverse wave velocity $c_T = 3122~~\textrm{mm/}\mu\textrm{s}$. The dispersion curves of the first propagating symmetrical Lamb and SH modes are shown in Fig.~\ref{fig2a}. For the sake of clarity, the SH and Lamb modes are labeled $SH_i$ and $S_i$, respectively. As highlighted by previous studies~\cite{meitzler1965backward}, over the frequency range from the ZGV resonance to the cut-off frequency, a right-going $S_{2}$ Lamb mode is associated with a negative wave number, \ie, a negative phase velocity $\omega/k$ and a positive group velocity $\partial \omega/\partial k$ resulting in a backward propagation. This mode is labelled $S_{2b}$ in order to differentiate it from the forward propagating $S_2$ mode.

Another crucial feature is the occurrence of a minimum for the $S_1$ and $S_{2b}$ dispersion branches at the same frequency $f = f_{ZGV}$ and opposite wave numbers $k = \pm k_{ZGV}$. This minimum corresponds to a zero-group velocity point, where the group velocity vanishes while the phase velocity remains finite~\cite{prada2005laser,holland2003air}.\\

Figure~\ref{fig2b} displays the frequency-dependent reflection coefficients for a normally incident $S_1$ wave on a free edge in the frequency band starting from $f\times\left(2h\right) = f_{ZGV} \times\left(2h\right) = 2.87~\textrm{MHz}\cdot\textrm{mm}$ to $f\times\left(2h\right) = 3.30~\textrm{MHz}\cdot\textrm{mm}$. At normal incidence, due to the decoupling between Lamb and SH modes, the incident mode is converted into a combination of Lamb modes only. A remarkable feature in Fig.~\ref{fig2b} is the perfect conversion from the forward $S_1$ mode into the backward $S_{2b}$ mode in the vicinity of the ZGV-point. It arises from the equality between the wave numbers of the incident forward $S_1$ mode and the reflected $S_{2b}$ mode. As a consequence, these two modes are associated with similar stress-displacement fields. The only difference lies in their opposite Poynting vector directions. Thus, the stress-free boundary condition at the edge of the plate can be satisfied with a simple combination of $S_1$ and $S_{2b}$ modes stress field, leading to a reflection coefficient $r_{S_1\vert S_{2b}} = -1$.\\

\begin{figure*}[!ht]
\centering
\begin{minipage}[c]{0.7\textwidth}
\subfigure{\includegraphics[width=\textwidth]{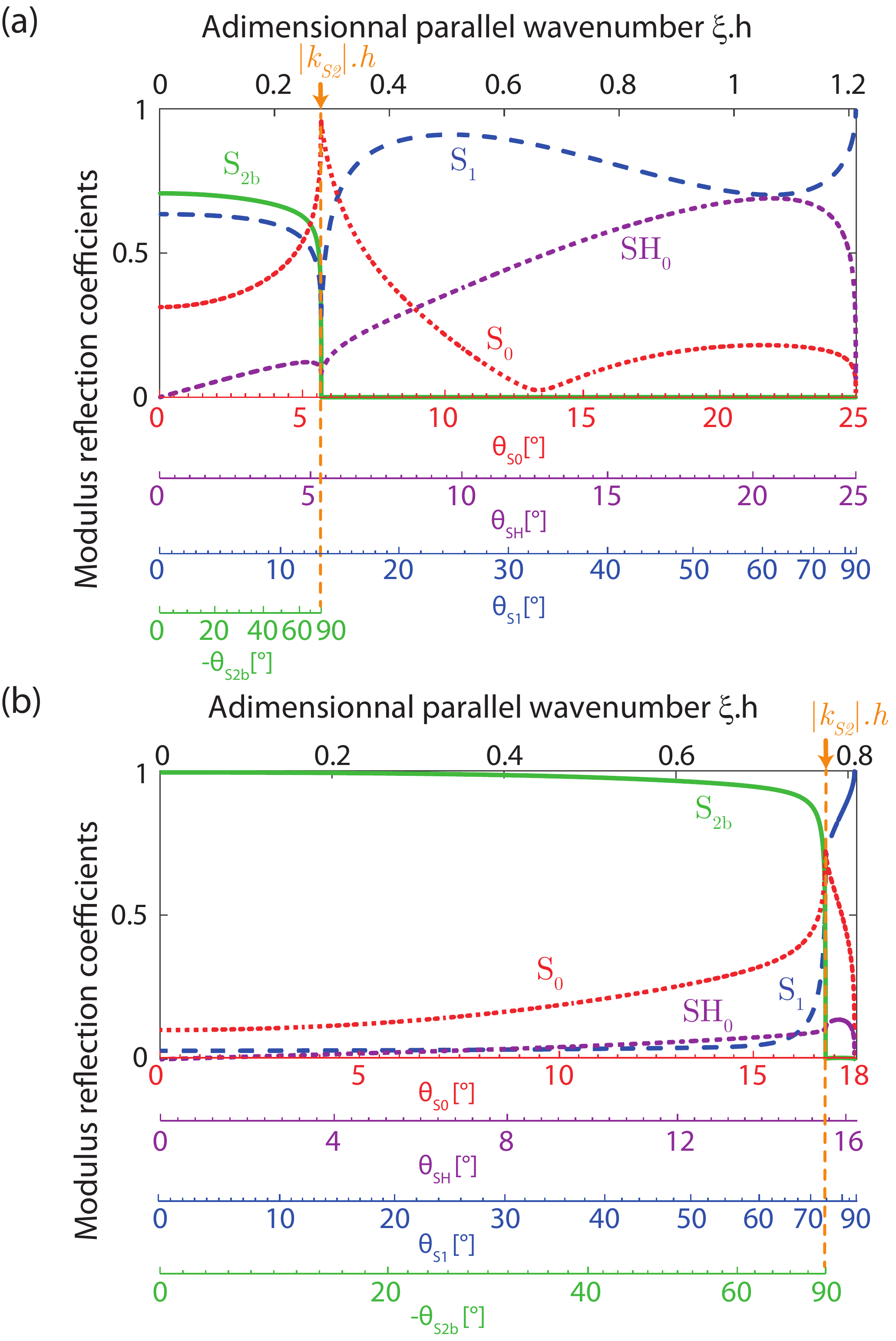}\label{fig3a}}
\subfigure{\label{fig3b}}
\end{minipage}\hfill
\begin{minipage}[c]{0.25\textwidth}
\caption{(color online) Modulus of the reflection coefficients of the propagating Lamb and SH modes for an obliquely incident $S_1$ Lamb waves as function of the dimensionless parallel wave number $\xi\cdot h$ for (a)~$f\times\left(2h\right) = 3.00~\textrm{MHz}\cdot\textrm{mm}$ and (b)~$f\times\left(2h\right)  = 2.87~\textrm{MHz}\cdot\textrm{mm}~\sim f_{ZGV}\times\left(2h\right) $. The corresponding angles of incidence and reflection are also indicated.}
\label{fig3}
\end{minipage}
\end{figure*}

Nevertheless, this picture is only valid under a normal incidence since it implies a perfect decoupling between the SH and Lamb modes. Figure~\ref{fig3a} shows the evolution of the reflection coefficients $\left| r_{S_1\vert n} \right|$ with the angle of incidence of the incoming $S_1$ Lamb mode at $f\times\left(2h\right)= 3.00~\textrm{MHz}\cdot\textrm{mm}$. Far from the ZGV-point, the $S_1$ mode is simultaneously reflected into $S_0$, $S_1$, $S_{2b}$ and $SH_0$ modes. As $\left| k_{S_{2b}} \right|  < \left| k_{S_{1}} \right|$, the conversion of $S_1$ into $S_{2b}$ only occurs below the critical angle $\theta_c = \arcsin{\left(\left| k_{S_{2b}} / k_{S_{1}} \right|\right)}$. Interestingly, this conversion remains almost constant for $\left| \theta_{S_{1}} \right| < \theta_c$. The $S_{2b}$ mode is thus excited over the whole angular spectrum. This feature holds close to the ZGV-point (see Fig.~\ref{fig3b}), moreover the conversion is now nearly perfect $\left(\left| r_{S_1\vert S_{2b}} \right| \sim 1 \right)$ over the whole angular spectrum $\left(\theta_c \rightarrow \pi/2\right)$. Besides, as $r_{S_1\vert S_{2b}} = r_{S_{2b}\vert S_{1}}$, the conversion of an incident $S_{2b}$ mode into $S_1$ is also nearly perfect.
Close to the ZGV resonance, the free edge thus acts as a nearly perfect negative reflecting mirror for both $S_1$ and $S_{2b}$ modes. This striking behavior is now investigated experimentally in the next section.

\section{Negative reflection of a cylindrical wave-field: Experimental results}\label{sec3}

The conversion between forward and backward modes at the free edge of a plate results in negative reflection. This effect is a direct consequence of their oppositely signed wave numbers and of the conservation of the parallel component of the wave vectors at the edge of the plate. In the first two parts, we will take advantage of negative reflection to experimentally demonstrate a tunable focusing mirror above the ZGV resonance. In a third part, the free edge will be shown to act as a phase conjugating mirror in the vicinity of the ZGV point.

\begin{figure}[!ht]
\centering
\subfigure{\includegraphics[width=\columnwidth]{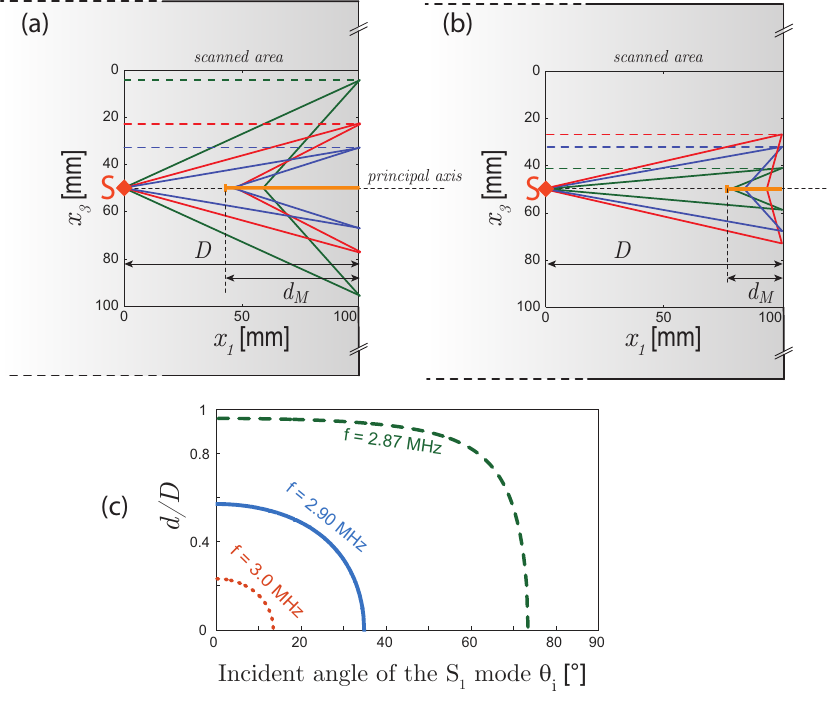}\label{fig4a}}
\subfigure{\label{fig4b}}
\subfigure{\label{fig4c}}
\caption{(color online) Network ray tracing of the reflection by a free edge for a plate of $2h=1~$mm at (a) $f = 2.90~\textrm{MHz}$ and (b) $f = 3.00~\textrm{MHz}$. The $S_{2b}$ reflected rays cross on the principal axis in the segment limited by the edge and the point at a distance $d_M$ (see Eq.~\ref{eq:d_D})\ from the edge (orange thick line). (c) Dependence of the ratio $d/D$ with the incident angle of the $S_1$ mode at frequencies $f = 2.87~\textrm{MHz}~\sim f_{ZGV}$ (green dashed line), $f = 2.90~\textrm{MHz}$~(blue plain line) and $f = 3.00~\textrm{MHz}$~(red dots).}
\label{fig4}
\end{figure}

\subsection{Experimental demonstration of the plate edge acting as a tunable focusing mirror for an incident $S_1$ Lamb mode}\label{sec3a}

The principle of the focusing plane mirror for an incident $S_1$ Lamb mode is depicted in Fig.~\ref{fig4} for a plate of thickness $2h = 1~$mm. Considering a point source $S$ emitting selectively the $S_1$ mode at $f = 2.90~$MHz [Fig.~\ref{fig4a}] and $f= 3.00~$MHz [Fig.~\ref{fig4b}], the network of the $S_{2b}$ reflected rays corresponding to different angles of incidence cross the principal axis at different distances $d$ from the free edge,
\begin{equation}
d = {D} \left|\frac{\tan{\theta_{i}}}{\tan{\theta_{S_{2b}}}}\right|.
\label{eq:d_D_tan}
\end{equation}
At the critical angle of incidence $\left( \theta_{S_{2b}} = \pi/2\right)$, the intersection with the principal axis always occurs on the edge. On the contrary, in the limit of small incident angles, it occurs at a maximal distance $d_M$ given by
\begin{equation}
d_M \simeq {D} \left|\frac{k_{S_{2b}}}{k_{S_1}}\right|,
\label{eq:d_D}
\end{equation}
where $D$ is the distance from the source to the edge. The non-linearity of $\theta_{S_{2b}}$ as a function of the incident angle $\theta_i$ [Eq.~\ref{eq:d_D_tan}] causes a non-uniform deposit of energy along the principal axis between the edge and the point located at a distance $d_M$. Fig.~\ref{fig4c} displays the dependence of $d/D$ with the incident angle $\theta_{i}$ at three frequencies. The concavity of the different curves at each frequency indicates that the density of rays increases gradually on the principal axis as the angle of incidence decreases. Thus, most energy is focused at a distance $d_M$ from the edge. This is all the more true as the frequency approaches $f_{ZGV}$.\\

In order to illustrate the focusing effect induced by the plate edge, the $S_1$ mode generation should be made as selective as possible. Experimentally, this is performed by means of an array of 64 transducers placed at a distance $D = 10~$cm from the free edge, as shown in Fig.~\ref{fig5} (see Appendix for details). The plate dimensions are chosen so that only the facing free edge causes reflection during the recording time of the experiment. The out-of-plane displacement is measured from the transducers' array to the plate edge with a heterodyne interferometer. The measurement is made over a grid of points that maps $10 \times 10~$cm$^2$ of the plate surface, with a pitch of $0.5~$mm. Signals detected by the optical probe are fed into a digital sampling oscilloscope and transferred to a computer. A spatio-temporal discrete Fourier transform (DFT) of the recorded wavefronts is then performed from $2.90$ to $3.10$ MHz for wave numbers ranging from $-4$ to $4$~mm$^{-1}$. Then, the experimental dispersion curves are obtained by angularly integrating the spatial Fourier plane at each frequency.\\

\begin{figure}[!ht]
\centering
\includegraphics[width=\columnwidth]{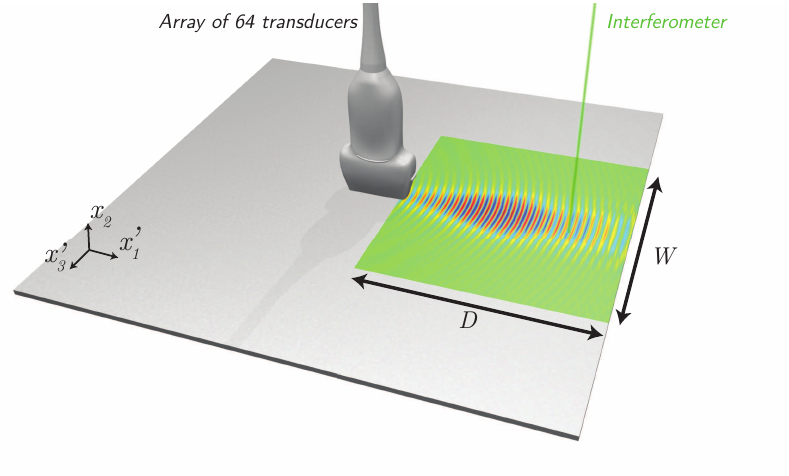}
\caption{(color online) Experimental setup. The $S_1$ mode is generated selectively in the Duralumin plate of thickness $2h =1~$mm by an array of 64 transducers. The out-of-plane component of the local vibration is measured with a heterodyne interferometer over a grid of point that maps $10 \times 10~$cm$^2$ of the plate surface. The grid pitch is 0.5 mm. }
\label{fig5}
\end{figure}

\begin{figure*}[!ht]
\centering
\begin{minipage}[c]{0.7\textwidth}
\subfigure{\includegraphics[width=\columnwidth]{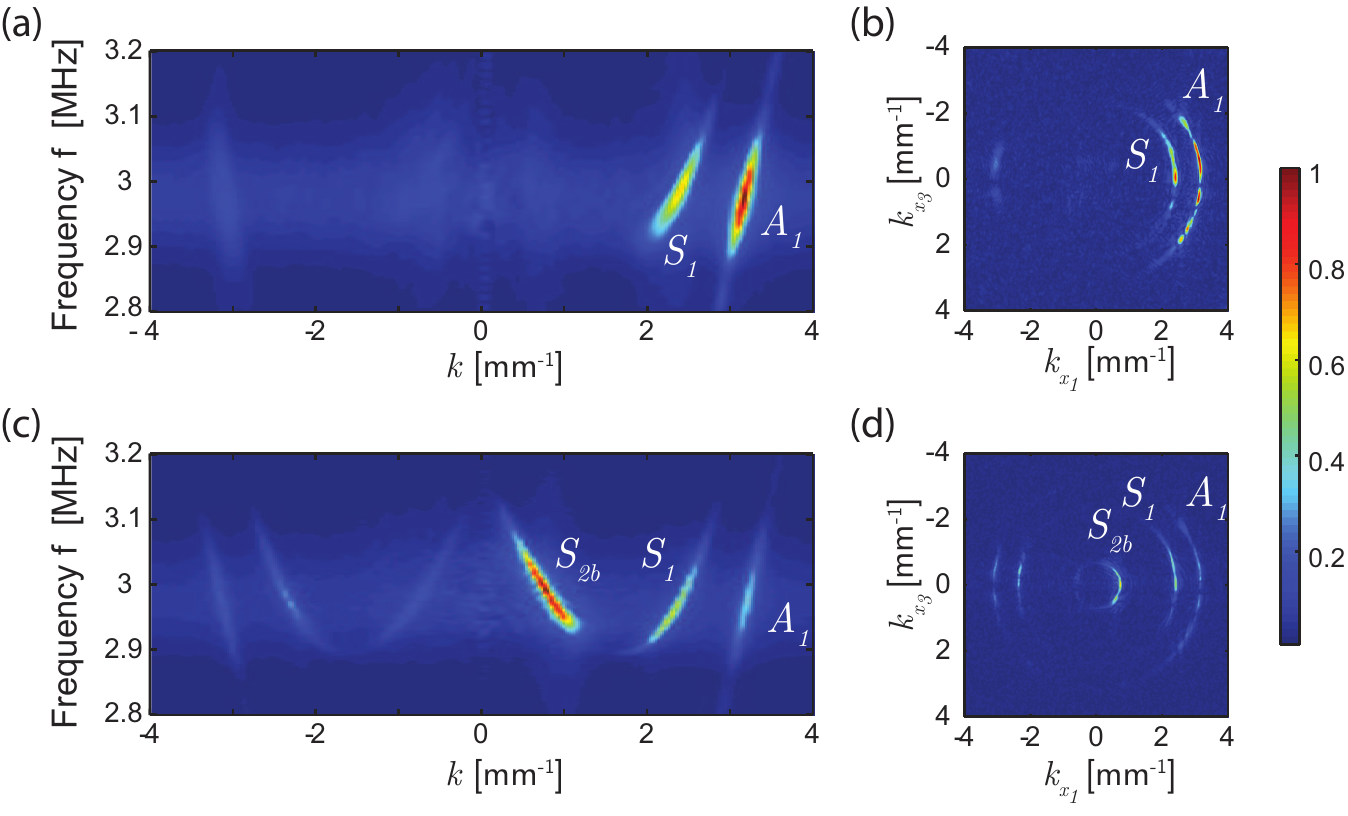}\label{fig6a}}
\subfigure{\label{fig6b}} \subfigure{\label{fig6c}} \subfigure{\label{fig6d}}
\end{minipage}\hfill
\begin{minipage}[c]{0.25\textwidth}
\caption{(color online) (a) Experimental dispersion curves obtained with signals truncated before the reflection at the free edge. The $S_1$ mode is generated preferentially along with the antisymmetric $A_1$ mode. (b) Incident wave-field in the spatial Fourier plane at $f = 3.00~$MHz. (c) Experimental dispersion curves obtained with signals including the reflection. (d) Recorded wave-field in the spatial Fourier plane at $f = 3.00~$MHz.}
\label{fig6}
\end{minipage}
\end{figure*}

To characterize the wave-field generated by the array of transducers, a dispersion curve is first computed over a time range of $70~\mu$s that excludes the echoes due to reflections on the free edge. The result is displayed in Fig.~\ref{fig6a}. It appears that the $S_1$ mode is efficiently generated, unlike the $S_{2b}$ mode. The antisymmetric $A_1$ mode is also significantly excited. Nevertheless, this mode is only reflected into anti-symmetric ones at the free edge for symmetry reasons. Hence, its excitation does not affect the analysis of symmetrical modes. The angular dependence of the $S_1$ mode excitation can be evaluated at each frequency by considering the spatial Fourier plane. Fig.~\ref{fig6b} highlights the anisotropy and angular inhomogeneities of our generation scheme at the frequency $f = 3.00~$MHz. This can be accounted for by the finite size of the source and the imperfect coupling between the transducers' array and the plate surface. Nevertheless, the angular range is sufficient to access the critical angle $\theta_c$, hence providing a reflection of the $S_{2b}$ mode over the whole angular spectrum.\\

Fig.~\ref{fig6c} depicts the dispersion curve computed over a time range from $0$ to $200~\mu$s that includes the waves reflected by the free edge. As expected theoretically, a significant part of the incident right-going $S_1$ mode is converted into a left-propagating $S_{2b}$ mode at the free edge, resulting in negative reflection. The left-going backward propagating $S_{2b}$ mode is also associated with a positive wave number $k$, because of its negative phase velocity. Figure~\ref{fig6d} shows the corresponding spatial Fourier plane at $f = 3.00$~MHz. It brings to light the wide angular distribution of the reflected $S_{2b}$ wave that will give rise to the focusing of the reflected wavefront.

\begin{figure*}[!ht]
\centering
\begin{minipage}[c]{0.65\textwidth}
\subfigure{\includegraphics[width=\columnwidth]{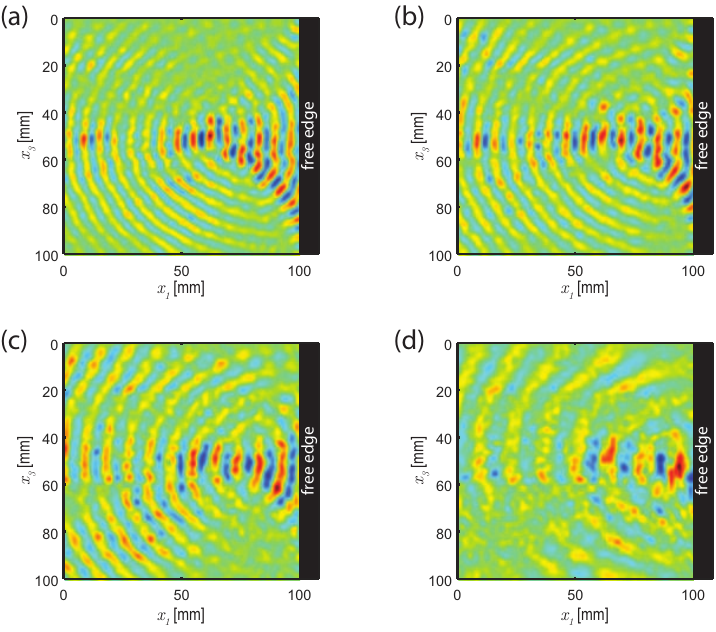}\label{fig7a}}
\subfigure{\label{fig7b}} \subfigure{\label{fig7c}} \subfigure{\label{fig7d}}
\end{minipage}\hfill
\begin{minipage}[c]{0.25\textwidth}
\caption{(color online) Normalized negatively reflected field for an incident $S_1$ mode at frequency $f = 2.95~$MHz (a), $f = 2.97~$MHz (b), $f = 3.00~$MHz (c) and $f = 3.05~$MHz (d). The focusing distance can be tuned by changing the frequency, \ie playing on the ratio between $k_{S1}$ and $k_{S_{2b}}$.}
\label{fig7}
\end{minipage}
\end{figure*}

To demonstrate this focusing, the $S_{2b}$ reflected wave-field is extracted from the data by filtering the Fourier plane at each frequency with a spatial low pass filter with a cut-off $k_c < k_{S_1}$. Fig.~\ref{fig7}(a)-(d) shows the reflected wave-field at different frequencies. Given the position of the source, the focusing distance from the free edge is determined by the ratio $\left| k_{S_{1}} \right| / \left| k_{S_{2b}} \right|$ as depicted in Eq.~(\ref{eq:d_D}). As this ratio increases with frequency, the focal spot goes closer to the free edge. In the limit of infinite wavelength for the mode $S_{2b}$ (\ie at its cut-off frequency), the focusing occurs directly on the free edge. To conclude, Fig.~\ref{fig7} is a nice illustration of the tunable focusing property that can exhibit a simple free edge thanks to negative reflection. 

\subsection{Experimental evidence of the plate edge acting as a tunable focusing mirror for an incident $S_{2b}$ Lamb mode}\label{sec3b}

\begin{figure*}[!ht]		
\centering
\subfigure{\includegraphics[width=0.7\textwidth]{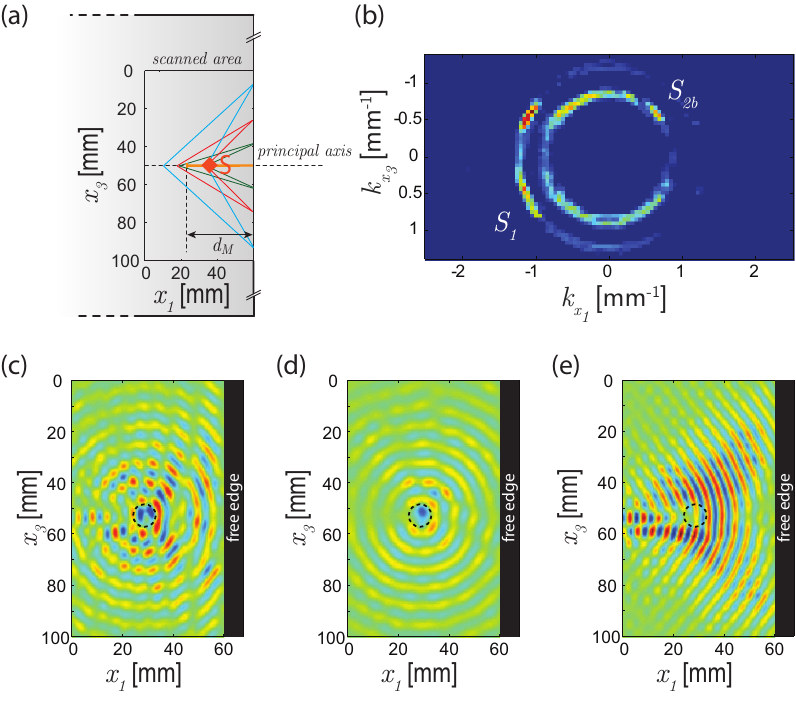}\label{fig8a}}	
\subfigure{\label{fig8b}}	\subfigure{\label{fig8c}}	
\subfigure{\label{fig8d}}	\subfigure{\label{fig8e}}	
\caption{(color online) Negative reflection of the $S_{2b}$ mode slihgtly above the ZGV point. (a) Network ray tracing of the reflection by a free edge for a plate of  a plate of thickness $2h = 1.5~$mm at $f\times\left(2h\right)= 2.86~\textrm{MHz}\cdot\textrm{mm}$ (b) Field spatial Fourier spectrum at $f\times\left(2h\right) =  2.86~\textrm{MHz}\cdot\textrm{mm}$. The backward $S_{2b}$ mode is generated selectively (inner circle) and is converted at the free edge in the $S_1$ mode. (c) Total wave-field resulting from the superposition of the backward $S_{2b}$ mode generated by the transducer and the forward $S_1$ mode converted on the free edge. (d) Low pass filtered wave-field corresponding to the generated backward $S_{2b}$-mode and (e) high-pass filtered wave-field corresponding to the the reflected forward-propagating $S_1$-mode.}
\label{fig8}
\end{figure*}

The principle of the focusing plane mirror is now studied for an incident $S_{2b}$ Lamb mode. Slightly above the ZGV point, the incident mode is expected to be exclusively reflected into the $S_1$ Lamb mode at the free edge of the plate. The principle of the focusing plane mirror for an incident $S_{2b}$ Lamb mode emitted selectively at $f\times\left(2h\right)= 2.86~\textrm{MHz}\cdot\textrm{mm}$ is depicted in Fig.~\ref{fig8a} for a plate of thickness $2h = 1.5~$mm. Contrary to the case of an incident $S_1$ lamb mode, as $\left| k_{S_{2b}} \right| < \left| k_{S_{1}} \right|$, the focusing occurs at a distance further than the distance of the source from the edge. In the limit of small angles, it occurs at a minimal distance $d_M$ as given by Eq.~(\ref{eq:d_D}), whereas, in the limit of $\theta_i \rightarrow \pi/2$, it occurs at an infinite distance from the edge. However, as in the previous part, the non-linearity of $\theta_{S1}$ as a function of $\theta_i$ causes most of the energy to be focused at a distance $d_M$ from the edge.\\

In this subsection, the transducers' array cannot be used as it would perturb the reflected wave-field. Hence, the source consists in a piezo-electric transducer of $7$~mm-diameter placed at a distance $D = 30$~mm from the edge of a Duralumin plate of thickness $2h~=~1.5~$mm. A $10~\mu$s chirp signal spanning the frequency range $1.8 - 2.0$~MHz is sent to the transducer which generates a cylindrical incident wavefront in the plate.  As in the previous experiment, the vibration of the plate is measured with a heterodyne interferometer over a grid of points that maps $10 \times 6~$cm$^2$ of the plate surface. The dimension of the plate is large enough to avoid reflections from the other edges during the recording time of the experiment.\\

A spatial DFT of the recorded wave-field is performed at $f\times\left(2h\right)= 2.86~\textrm{MHz}\cdot\textrm{mm}$, over a time range from $0$ to $250~\mu$s [see Fig.\ref{fig8b}]. In the spatial Fourier plane, the inner circle accounts for the cylindrical generation of the backward $S_{2b}$ mode, whereas the contribution from the $S_1$ mode (outer circle) only occurs for negative $k_{x_1}$ values, meaning it only originates from the reflection on the free edge. The selective excitation of the $S_{2b}$-mode is explained by the matching of its wavelength with the transducer diameter. Contrary to the previous experiment, as $\left| k_{S_{2b}} \right| < \left| k_{S_{1}} \right|$, the reflection is angularly limited to $\theta_c = \arcsin{\left(\left|k_{S_{2b}}/k_{S_{1}}\right|\right)}$, resulting in an arc of a circle for the reflected $S_1$ mode in the spatial Fourier plane. The corresponding wave-field is investigated in Fig.~\ref{fig8}(c)-(e). Fig.~\ref{fig8c} displays the wave-field resulting from the superposition of the backward $S_{2b}$ mode generated by the transducer and the forward $S_1$ mode reflected by the free edge. In order to clarify our interpretation of data, the contribution from each mode can be extracted by spatial filtering in the Fourier plane. Fig.~\ref{fig8d} represents the incident $S_{2b}$ mode whose excitation is almost perfectly isotropic, as expected from its circular distribution in the Fourier plane [Fig.~\ref{fig8b}]. The reflected wave-field is shown in Fig.~\ref{fig8e}. As expected from Eq.~(\ref{eq:d_D_tan}), the focusing occurs at a distance greater than $D$. Moreover, as the $S_1$ mode is reflected over a limited angular range [see Fig.~\ref{fig8b}], the focusing operation is not as efficient as in Sec.~\ref{sec3a}. The reflected wave-field is also disturbed by the presence of the transducer whose position is represented by a black dashed line in Figs.~\ref{fig8}(c)-(e).  At last, it is important to note that, whereas, in the first experiment, all the reflected rays were of finite length [Fig.~\ref{fig4}], paths of infinite length ($\theta_i \rightarrow  \pi/2$) here cannot be collected during the recording time of our measurements.

\subsection{Experimental demonstration of the plate edge acting as phase conjugating mirror}\label{sec3c}

As shown in Sec.~\ref{sec2}, the conversion between the $S_1$ and $S_{2b}$ modes at a free edge is almost perfect in the vicinity of the ZGV point. Moreover, for a point-like source, the negatively reflected wave-field should focus back exactly at the initial source location (see Fig.~\ref{fig4}). Hence the free edge should act as a phase conjugating mirror near the ZGV resonance. The aim of this subsection is to investigate experimentally this striking behavior.\\

\begin{figure*}[!ht]		
\centering
\subfigure{\includegraphics[width=0.5\textwidth]{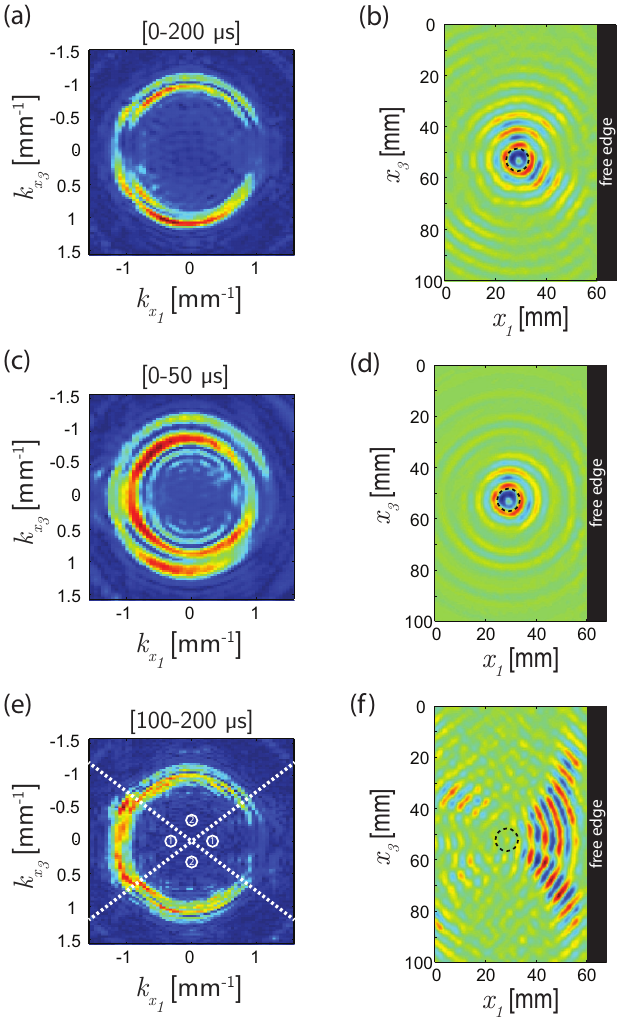}\label{fig9a}}	
\subfigure{\label{fig9b}}
\subfigure{\label{fig9c}}
\subfigure{\label{fig9d}}
\subfigure{\label{fig9e}}
\subfigure{\label{fig9f}}
\caption{(color online) Negative reflection in the vicinity of the ZGV point. (a),~Spatial Fourier spectrum of the total wave-field computed at the frequency-thickness product $f\times\left(2h\right)= 2.85~\textrm{MHz}\cdot\textrm{mm}$ over the first $200~\mu$s. (b),~Total wave-field in the real space deduced from an inverse DFT of the spatial Fourier spectrum displayed in (a). (c),~Spatial Fourier spectrum of the incident wave-field computed at the frequency-thickness product $f\times\left(2h\right)= 2.85~\textrm{MHz}\cdot\textrm{mm}$ over the first $50~\mu$s. (d), Incident wave-field in the real space deduced from an inverse DFT of the spatial Fourier spectrum displayed in (c). (e),~Spatial Fourier spectrum of the reflected wave-field computed at the frequency-thickness product $f\times\left(2h\right)= 2.85~\textrm{MHz}\cdot\textrm{mm}$ over the time range $\left[ 100 - 200 \right]~\mu$s. (f),~Reflected wave-field in the real space deduced from an inverse DFT of the spatial Fourier spectrum over the angular range labelled (1) in (e). }
\label{fig9}
\end{figure*}

The experimental configuration is described in the previous subsection. The wave is generated at the frequency-thickness product $f\times\left(2h\right)= 2.85~\textrm{MHz}\cdot\textrm{mm}$, assimilated to the ZGV point. A spatial DFT of the recorded wave-field is performed over the first $200~\mu$s including the waves reflected by the free edge. The associated spatial Fourrier transform is displayed in Fig.~\ref{fig9a}. Two circles associated with close $k$ values correspond to the forward $S_1$ (outer circle) and the backward $S_{2b}$ (inner circle) modes. An inverse DFT yields the corresponding wave-field in the real space [see Fig.~\ref{fig9b}]. Contrary to the wave-field measured above the ZGV resonance [Fig.~\ref{fig8c}], the cylindrical incident and reflected wave-fronts are here impossible to discriminate as they exhibit the same curvature.\\

To extract the contribution of the incident wave-field, a spatial DFT is performed on the first $50~\mu$s. This time window excludes the reflection on the free edge of most  slow modes [see Fig.~\ref{fig9c}]. As $\lambda_{S_1} \simeq \lambda_{S_{2b}}$ in the vicinity of the ZGV resonance, the generation cannot be selective and both modes are excited. However, it appears that the $S_{2b}$ mode is still preferentially and isotropically generated than the $S_1$ mode. The corresponding wave-field is displayed in the real space in Fig.~\ref{fig9d}. As expected from its circular distribution in the Fourier plane, the incident wave-field is almost perfectly isotropic.\\

The reflected contribution is now isolated by computing the DFT of the recorded wave-field over a time window from $100$ to $200~\mu$s [see Fig.~\ref{fig9e}]. However, because of their near zero group velocities, the incident and reflected waves cannot be perfectly separated. The angular range labelled (1) in  Fig.~\ref{fig9e} (small angles of reflection) corresponds to the reflected wave-field whereas the angular range labelled (2) in  Fig.~\ref{fig9e} (large angles) corresponds to the residual incident wave-front. Fig.~\ref{fig9e} shows that, in agreement with our theoretical calculations, the incident $S_{2b}$ mode is efficiently reflected into the right-going $S_1$ mode $\left( k_{x_1} < 0 \right)$. An inverse DFT of the signal in the angular range labelled (1) in Fig.~\ref{fig9e} yields the reflected wave-field in the real space [see Fig.~\ref{fig9f}]. As expected, the reflected wave focuses back on the source location whose position is represented by a black dashed circle in  Fig.~\ref{fig9f}. The free edge is shown to act as a phase conjugating mirror. Note that the presence of the transducer pertubs the reflected wave-field and prevents us from observing its propagation at later times.\\

This experiment demonstrates that the free edge of the plate acts as as a passive phase conjugating mirror in the vicinity of the ZGV resonance. The negative reflection phenomenon thus offers numerous possibilities to control the propagation of Lamb waves in the vicinity of a free edge.

\section{Conclusion}

In summary, we have studied semi-analytically and experimentally the interaction of Lamb waves with the free edge of a plate. In particular, the conversion between forward and backward Lamb modes, that gives rise to negative reflection, has been investigated in details. Interestingly, it has been theoretically shown that this conversion is uniform over most of the angular spectrum. Experimentally, this has been taken advantage of to show how a simple free edge can act as an efficient tunable focusing mirror. Last but not least, close to the ZGV resonance, the conversion between forward and backward modes has been shown to be total and the plate edge shown experimentally to act as a phase conjugating mirror. This striking behavior opens new perspectives for the manipulation of Lamb waves. In particular, negative reflection gives rise to enhanced back-scattering that can be taken advantage of for the detection of defects in non destructive testing. Besides, an association of the phase conjugation phenomenon with the sensitivity of the ZGV resonance~\cite{prada2005laser} can be fruitful and may lead to a new type of acoustic sensors.

\section{Acknowledgments}

The authors wish to thank T. W. Murray for drawing our attention to the topic and V. Pagneux for fruitful discussions. The authors are grateful for funding provided by LABEX WIFI (Laboratory of Excellence within the French Program Investments for the Future, ANR-10-LABX-24 and ANR-10-IDEX-0001-02 PSL*) and by the Agence Nationale de la Recherche (ANR-15-CE24-0014-01, Research Project COPPOLA). B.G. acknowledges financial support from the French ``Direction Générale de l'Armement''(DGA).

\section*{Appendix A : Numerical method}\label{AnnexeA}

The numerical method presented in this appendix is adapted from the one implemented by Pagneux~\cite{pagneux2006revisiting}. The elasticity equations (\ref{eq:elasticity_v1}) and (\ref{eq:elasticity_v2}) can be written 
\begin{equation}
\partial_{x_1}{\left( \begin{array}{c}	\mathbf{X}\\	\mathbf{Y}\\	\mathbf{Z}	\end{array} \right)} = \left( \begin{array}{ccc} 0& \mathbf{F} &0\\\mathbf{G}& 0 &0\\0& 0 &\mathbf{H}\end{array}\right) \cdot \left( \begin{array}{c}	\mathbf{X}\\	\mathbf{Y}\\	\mathbf{Z}	\end{array}
\right),
\label{eq:elast_mat}
\end{equation}
where $\mathbf{X} = \left(u_1,\sigma_{12} \right)^T$, $\mathbf{Y} = \left(-u_2,\sigma_{11} \right)^T$, $\mathbf{Z} = \left(u_3,\sigma_{13} \right)^T$ and where $F$, $G$ and $H$ are the operators,
\begin{equation}
F =\left(
\begin{array}{cc} 
-\left(\frac{1}{\gamma}\right)                       & -\left(\frac{\gamma-2}{\gamma}\right)\partial_y\\
\left(\frac{\gamma-2}{\gamma}\right)\partial_y\quad  & -\left[\Omega^2-4\left(\frac{\gamma-1}{\gamma}\right)\partial_{y^2}\right] 
\end{array}\right),
\label{eq:F}
\end{equation}
\begin{equation}
G =\left( \begin{array}{cc} \Omega^2& \partial_y\\
   -\partial_y& 1 \end{array}\right),
\label{eq:G}
\end{equation}
\begin{equation}
H =\left( \begin{array}{cc} 0 & \Omega^2-\partial_{y^2}\\ 
1 & 0 \end{array}\right).
\label{eq:H}
\end{equation}
The boundary condition expressed in Eq.~(\ref{eq:elasticity_v4}) can be expressed in term of $\mathbf{Y}$ by writing 
\begin{equation}
\sigma_{12}\left(\mathbf{Y}\right) = \left (\frac{\gamma -2}{\gamma}\right )\sigma_{11} + 4\left (\frac{\gamma-1}{\gamma}\right )\partial_{x_1}{u_2}.
\label{eq:boundary_r_Y}
\end{equation}
Then, the whole problem as the boundary conditions can be expressed in term of $\mathbf{X}$, $\mathbf{Y}$ and $\mathbf{Z}$. The problem is solved numerically by using the MATLAB Differentiation Suite developed by Weideman and Reddy~\cite{weideman2000matlab}. This suite can be used to solve ordinary differential equations using a Chebyshev collocation spectral method. Equations are spatially discretized along the $x_2$ axis.
Any function $q_N(x_2)$ on the domain $\left[ -1,1\right]$ may also be approximated as 
$q_N(x_2) = \sum_{k=1}^N \hat{q}_k T_k(x_2)$, where, $\hat{q}_k=q\left(x_{2}^{\left(k\right)}\right)$, the functions $T_k(x_2)$ are the Chebyshev polynomials of order $k$ and the interpolation points are the Chebyshev-Gauss-Labatto points 
$x_{2}^{\left(k\right)} = \cos{\left(k\pi/(N-1)\right)}$, for $k=0,\ldots ,N-1$. Then, the spatial derivatives  of $q_N\left(x_2\right)$ along the $x_2$ coordinate can be evaluated at the collocation points by the matrix multiplication with the differentiation matrices $D_N$. To solve our problem, the displacement and stress tensor components are written respectively as
\begin{eqnarray}
u_i &=& \sum_{k=1}^N \hat{u}_{i}^{\left(k\right)} T_k{\left(x_2\right)}, \label{u_i_app} \\ 
\sigma_{ij} &=& \sum_{k=1}^N \hat{\sigma}_{ij}^{\left(k\right)} T_k{\left(x_2\right)}. \label{sigma_i_app}
\end{eqnarray}
The implementation of the boundary conditions $\boldsymbol{\sigma} \cdot \mathbf{n_2} = \mathbf{0}$ has to be performed carefully. The stress-free condition $\hat{\sigma}_{12}\left(x_2 = \pm 1\right)=0$ leads to $\hat{\sigma}_{12}^{\left(1\right)} = \hat{\sigma}_{12}^{\left(N\right)} = 0$. It can also be implemented by considering $\hat{\sigma}_{12}$ as a vector of dimension $N-2$. The boundary condition $\hat{\sigma}_{22}\left(x_2 = \pm 1\right)$ is implemented with the use of Eq.~(\ref{eq:boundary_r_Y}), leading to the direct relation between $\hat{\sigma}_{12}^{\left(i\right)}$ and $\hat{u}_{2}^{\left(i\right)}$,  for $i=1$ and $i=N-1$, $\hat{\sigma}_{11}^{\left(i\right)} = -4[(\gamma-1)/(\gamma-2)] \mathbf{l_i^T\mathbf{\hat{u}_2}}$, where $\mathbf{l_i}$ is the vector for the $i^\text{th}$ row of the differentiation matrix $\mathbf{D_1}$. The stress-free condition $\sigma_{13}\left(x_2 = \pm 1\right)=0$ is implemented by calculating the corresponding differentiation matrix $\mathbf{\tilde{D}_2}$  incorporating Robin conditions (see Ref.~\cite{weideman2000matlab} for details). With this discretization, Eqs. (\ref{u_i_app}) and (\ref{sigma_i_app}) become
\begin{equation}
\partial_{x_1}{\mathbf{X}} = \left( \begin{array}{cc} 
-\left(\frac{1}{\gamma}\right)\mathbf{I}          &  \mathbf{M_1} \\  
\left(\frac{\gamma-2}{\gamma}\right)\mathbf{D_1}  &  \mathbf{M_2} 
\end{array}\right) \cdot \mathbf{Y},
\label{eq:ODE_X_Y}
\end{equation}
\begin{equation}
\partial_{x_1}{\mathbf{Y}} = \left( \begin{array}{cc} 
\Omega^2 \mathbf{I}  &  \mathbf{D_1} \\
 - \mathbf{D_1}      &  \mathbf{I} 
\end{array}\right) \cdot \mathbf{X},
\label{eq:ODE_Y_X}
\end{equation}
\begin{equation}
\partial_{x_1}{ \mathbf{Z}} = \left( \begin{array}{cc} 
\mathbf{0} & -\left(\mathbf{\Omega^2I-\tilde{D}_2}\right) \\
\mathbf{I} & \mathbf{0} 
\end{array}\right) \cdot \mathbf{Z},
\label{eq:ODE_Z_Z}
\end{equation}
with
\begin{equation}
\mathbf{M_1} = - \left (\frac{\gamma-2}{\gamma}\right )\mathbf{D_1} - \left (\frac{4}{\gamma}\right ) \left (\frac{\gamma-1}{\gamma-2}\right ) \left(\begin{array}{c}  \mathbf{l_1}^T \\  
\mathbf{0} \\ 
\mathbf{l_N} \end{array}\right),
\label{eq:ODE_M_1}
\end{equation}
\begin{equation}
\mathbf{M_2} = - \Omega^2 \mathbf{I} -4 \left (\frac{\gamma-1}{\gamma}\right ) \mathbf{D_2} + 4\left (\frac{\gamma-1}{\gamma}\right ) 
\left(\mathbf{c_1 l_1^T}+\mathbf{c_N l_N^T}\right),
\label{eq:ODE_M_2}
\end{equation}
where $\mathbf{c_1}$ (respectively $\mathbf{c_N}$) is the first (respectively the last) column of the differentiation matrix $\mathbf{D_1}$ and $\mathbf{I}$ is the identity matrix. Each matrix has to be understood as having the dimension in agreement with the vectors it links.\\

Considering $N = 2N_e$ an even number, the problem can be reduced to the symmetric (respectively to the antisymmetric) problem by imposing the corresponding parity of the different displacement and stress tensor components. For example, in case of a symmetric problem, $u_1$, $u_3$, $\sigma_{11}$ and $\sigma_{13}$ are even functions whereas $u_2$ and $\sigma_{12}$ are odd functions. The diagonalization of this system of ordinary differential equations yields $\left(6N_e-2\right)$ eigenvalues and eigenvectors, corresponding to the $4N_e-2$ Lamb modes and $2N_e$ SH modes, which halves correspond to right-going modes and left-going modes. Considering an incident propagating right-going Lamb mode with an angle $\theta_i$ with respect to the normal direction, one can define the reflection angles $\theta_n$ corresponding to the different left-going modes, associated with the wave numbers $-k_n$ using Eq.~(\ref{eq:angle_reflection}). The strain-displacement field associated with each reflected mode is deduced by applying a rotation operator to the corresponding eigenvector (Eqs.(\ref{u_rotation})-(\ref{sigma_rotation}))

\begin{equation}
\mathbf{u^{\prime}} = \left( \begin{array}{c}u_1 \cdot \cos{\left(\theta\right)} - u_3 \cdot \sin{\left(\theta\right)} \\ u_2 \\ u_1 \cdot \sin{\left(\theta\right)} + u_3 \cdot \cos{\left(\theta\right)} \end{array}\right), 
\label{u_prime}
\end{equation}

\begin{widetext}
\begin{equation}
\boldsymbol{\sigma^{\prime}}\cdot\mathbf{n_1} = \left( \begin{array}{c} \sigma_{11} \cdot \cos^2{\left(\theta\right)} + \sigma_{33} \cdot \cos^2{\left(\theta\right)} - 2\cdot\sigma_{13}\cdot  \cos{\left(\theta\right)}\sin{\left(\theta\right)}\\
\sigma_{12} \cdot\cos{\left(\theta\right)} -\sigma_{23} \cdot\sin{\left(\theta\right)} \\ 
\left(\sigma_{11}-\sigma_{33}\right) \cdot\cos{\left(\theta\right)}\sin{\left(\theta\right)} +\sigma_{13}\cdot \left(\cos^2{\left(\theta\right)-\sin^2{\left(\theta\right)}} \right) \end{array}\right),
\label{sigma_prime}
\end{equation}
\end{widetext}

where $\sigma_{33}$ can be deduced from $\mathbf{X}$, $\sigma_{33}\left(\mathbf{X}\right) = \left(\sigma_{11} + 2 \cdot \mathbf{D_1}\cdot u_2\right)[(\gamma-2)/\gamma]$. Finally, the complex reflection coefficient can be computed using Eq.~(\ref{eq:boundary_condition}). Considering an obliquely incident right-going Lamb mode, there are $\left( 3 \cdot N_e -1\right)$ coefficients to determine the $\left( 2 \cdot N_e -1\right)$ left-going Lamb modes and the $N_e$ left-going SH modes. These coefficients satisfy $\left(3\cdot N_e -1\right)$ equations derived from the stress-free condition at the plate edge (Eq.~(\ref{eq:boundary_condition})). The cancellation of $\sigma_{11}^{\prime}$, $\sigma_{12}^{\prime}$ and  $\sigma_{13}^{\prime}$ gives $N_e$, $N_e-1$ and $N_e$ independent conditions, respectively. The computation is done with $N_e=100$.

\section*{Appendix B : Selective generation of the S1 mode}

\begin{figure}[!ht]
\centering
\includegraphics[width=0.6\columnwidth]{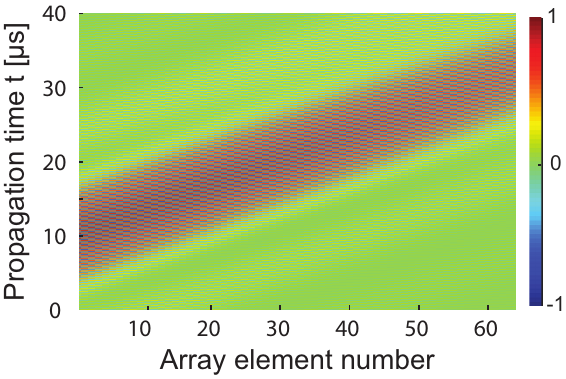}
\caption{(color online) B-Scan corresponding to the spatio-temporal excitation by the array of transducers. The array element number 64 is the element that is closest to the free edge.}
\label{appendix_B}
\end{figure}

The selective generation of the $S_1$ mode is achieved by using an array of 64 transducers. The transmitted signals are calculated using a 2D inverse DFT of the selected part of the dispersion curve $k\left(f\right)$, priorly convoluted with a Gaussian structuring element. The $S_1$ mode was selected between $2.9$ and $3.1$ MHz. The transducer array is driven by a Lecoeur Electronics 64-channel programmable device which allows for an $80$ MHz sampling frequency. The B-Scan corresponding to the spatio-temporal excitation by the array of transducers is displayed in Fig.~\ref{appendix_B}. The array element number 64 is the closest to the free edge.

\bibliographystyle{custom}

\end{document}